\documentclass[5p,number,sort&compress,times]{elsarticle}

\usepackage[utf8]{inputenc}
\usepackage{amssymb}
\usepackage{amsmath}
\usepackage{graphicx}
\usepackage{hyperref}
\usepackage{lineno}
\usepackage{pgf}
\usepackage{siunitx}
\usepackage{textcomp}
\usepackage{xspace}

\newcommand{\Kalman}{K\'alm\'an\xspace}

\renewcommand{\d}{\ensuremath{\operatorname{d}\!}}

\begin{document}
\begin{frontmatter}
\journal{Nuclear Instruments and Methods A}
\title{A New Three-Dimensional Track Fit with Multiple Scattering}
\author[A,B]{Niklaus Berger}
\author[A,B]{Alexandr Kozlinskiy}
\author[A]{Moritz Kiehn}
\author[A]{Andr\'e Sch\"oning\corref{mycorrespondingauthor}}
\cortext[mycorrespondingauthor]{Corresponding author}
\address[A]{Physikalisches Institut, Heidelberg University, Heidelberg, Germany}
\address[B]{now at Institut f\"ur Kernphysik and PRISMA cluster of excellence, Mainz University, Mainz, Germany}

\begin{abstract}
Modern semiconductor detectors allow for charged particle tracking with ever
increasing position resolution.
Due to the reduction of the spatial hit uncertainties,
multiple Coulomb scattering in the detector layers becomes the
dominant source for tracking uncertainties.
In this case long distance effects can be ignored for the momentum measurement, and the track fit  
can consequently be formulated as a sum of independent fits to hit triplets. 
In this paper we present an analytical solution for a three-dimensional 
triplet(s) fit in a homogeneous magnetic field based on a multiple scattering
model. 
Track fitting of hit triplets is performed using a linearization ansatz.
The momentum resolution is discussed for a typical spectrometer setup.
Furthermore the track fit is compared with other track fits for two 
different pixel detector geometries, namely the Mu3e experiment at PSI and a 
typical high-energy collider experiment.
For a large momentum range the triplets fit provides a significantly better
performance than a single helix fit. 
The triplets fit is fast and can easily be parallelized, which makes it ideal
for the implementation on parallel computing architectures.
\end{abstract}
\end{frontmatter}


\section{Motivation}

The trajectory of a free charged particle in a homogeneous magnetic field
is described by a helix.  The non-linear nature of the helix makes the
reconstruction of the three-dimensional trajectory from tracking
detector hits one of the main computational challenges in particle
physics.  To simplify the problem, the reconstruction is often
factorized into a two-dimensional circle fit in the plane transverse
to the magnetic field and a two-dimensional straight line fit in the
longitudinal plane\footnote{In the right-handed coordinate system we
  define the B-field orientation along the $z$-axis; the azimuthal
  angle $\varphi$ is defined in the transverse $x$-$y$ plane and the
  polar angle $\vartheta$ is defined in the longitudinal $z$-$s$ plane
  where $s$ is the track length parameter.}.  A non-iterative solution
to this problem was given by Karim\"aki \cite{Karimaki:1991:ecf}.  This
simplified treatment however does not make full use of the available
detector information and ignores correlations between the two planes,
which can be large especially for small helix
radii (low momentum particles) at small (large) polar angles $\vartheta \approx 0~ (\pi)$.

A further complication of the track reconstruction problem is the
treatment of multiple Coulomb scattering (MS) in the detector material,
which introduces correlations between the measurement points.  This
problem is addressed by \Kalman filters \cite{Kalman:1960:nal,
  Fruhwirth:1987:akf, BilloirEtAl:1990:spr} and broken line fits
\cite{Blobel:2006:nft, BlobelEtAl:2011:fac, Kleinwort:2012:gbl} which
 both give a correct description of the track parameter error matrix.
The methods however require computationally expensive matrix inversions
and potentially multiple passes.

In modern semiconductor pixel trackers, extremely precise
three-dimensional position information is available and tracking
uncertainties are dominated by MS except at the very
highest momenta.  Usually most of the material causing the scattering
is located in the sensors or very close to them (services, cooling,
mechanics etc.); therefore the scattering planes usually coincide with
the detection planes.  This is our motivation for developing a new
three-dimensional helix fit which treats MS in the
detector as the only uncertainty.  The resulting algorithm is based on
triplets of hits which can be fit in parallel.  The final result is
then obtained by combining all triplets.  The algorithm is
computationally efficient and well suited for track finding.
The first application of the algorithm is the all-pixel silicon
tracker \cite{BergerEtAl:2013:tme} of the Mu3e experiment \cite{mu3e-rp}.

\section{Triplet Track Fit}

\begin{figure}[t!]
\centering 
\includegraphics[width=0.45\textwidth]{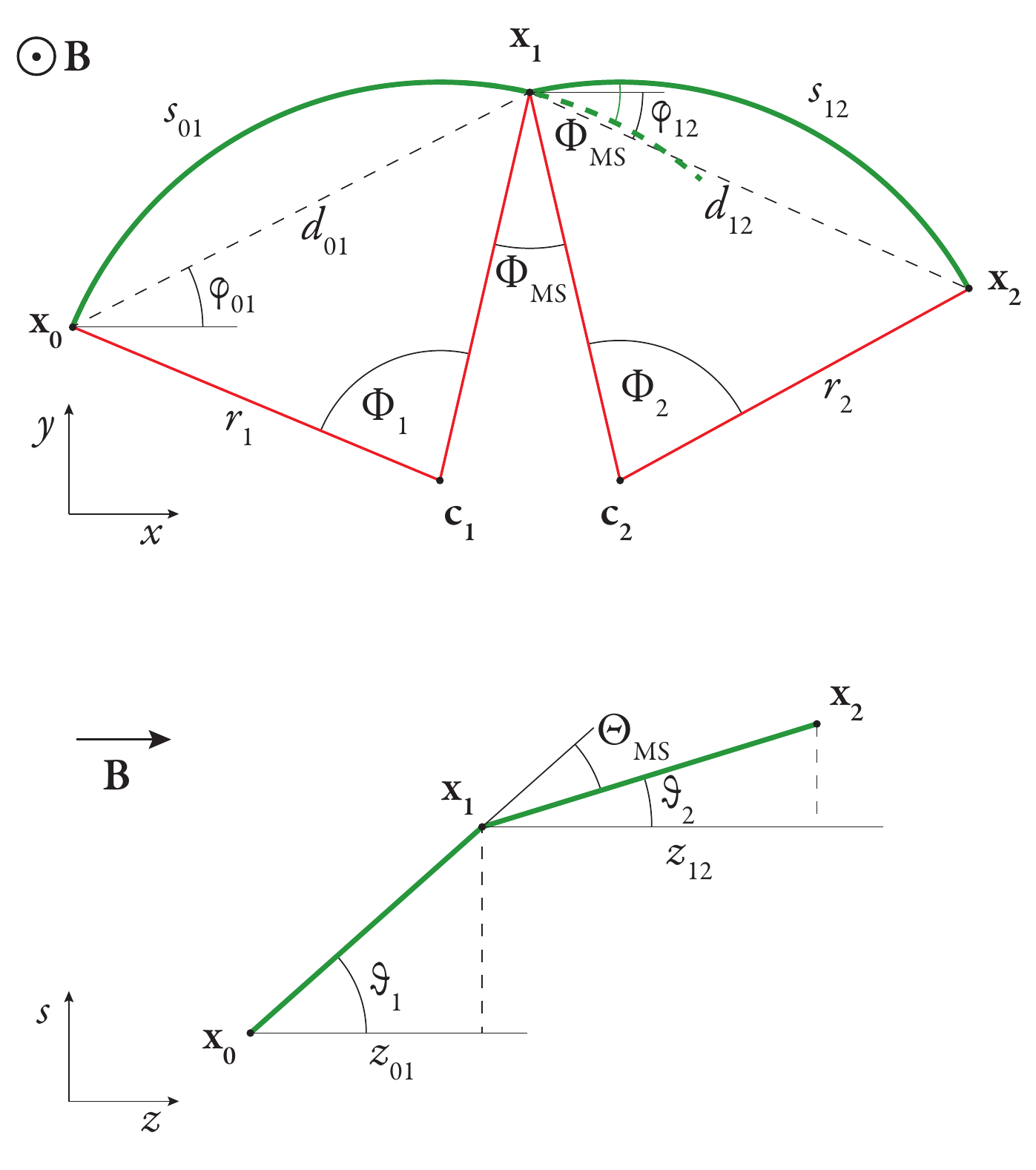}
\caption{Particle trajectory in a homogeneous magnetic field defined
  by a triplet of hits $\mathbf{x_0}$, $\mathbf{x_1}$ and
  $\mathbf{x_2}$, with particle scattering at $\mathbf{x_1}$.  
  The top view shows
  the projection to the plane transverse to the magnetic field,
  whereas the bottom view is a projection to the field axis-arclength
  ($s$) plane.  $r_1$ and $r_2$ are the transverse track radii before
  and after the scattering process, $s_{01}$ and $s_{12}$ the
  corresponding arclengths and $\Phi_{1}$ and $\Phi_{2}$ the bending
  angles.  $d_{01}$ and $d_{12}$ denote the transverse distances
  between $\mathbf{x_0}$ and $\mathbf{x_1}$, and $\mathbf{x_1}$ and
  $\mathbf{x_2}$, respectively.  The azimuthal angles of the
  corresponding distance vectors are labeled $\varphi_{01}$ and
  $\varphi_{12}$ and $\Phi_{MS}$ is the transverse scattering angle.
  In the longitudinal plane, $z_{01}$ and $z_{02}$ denote the
  distances between the measurement points along the field axis,
  $\vartheta_1$ and $\vartheta_2$ are the polar angles of the arcs and
  $\Theta_{MS}$ is the longitudinal scattering angle.}
\label{fig:triplet}
\end{figure}

The basic unit of the track fit is a triplet of hits in successive
detector layers.  In the absence of MS and energy losses, the
description of a helix through three points requires eight parameters,
namely a starting point (three parameters), an initial direction (two
parameters), the curvature (one parameter) and the distances to the
second and third point (two parameters).  
MS in the central plane requires two additional parameters to describe the
change in track direction\footnote{Two more parameters, describing a possible  
position offset at the central plane due to MS inside the material, can be ignored
for typical silicon trackers, where the sensor thicknesses 
are much smaller than the distances between the detector layers.}.
Three space points, which we assume to be measured without
uncertainties, do however only provide a total of nine coordinates;
additional constraints are thus needed to obtain the track parameters
and scattering angles.  These constraints can be obtained from
MS theory since the scattering angles depend statistically
on the particle type and momentum, and the material of the detector.

Starting from a hit triplet, see Figure~\ref{fig:triplet}, a
trajectory consisting of two arcs connecting the three-dimensional
space points is constructed.  It is assumed that the middle point
$\mathbf{x_1}$ lies in a scattering plane which deflects the particle
and thus creates a kink in the trajectory.  The corresponding
scattering angles in the transverse and longitudinal plane are denoted
by $\Phi_{MS}$ and $\Theta_{MS}$ respectively.

We assume that the particle momentum (and thus its three-dimensional
radius $R_{3D}$) is conserved\footnote{Energy loss due to ionization is
  usually small and can be either neglected or corrected for.}.  
The scattering angles $\Phi_{MS}$
and $\Theta_{MS}$ have a mean of zero and variances $\sigma_\vartheta^2 = \sigma_{MS}^2$
and $\sigma_\phi^2 = \sigma_{MS}^2 / \sin^2\vartheta$,
which can be calculated from MS theory, using e.g.~the Highland
approximation \cite{Highland:1975:prm}.  The task is thus to find a
unique $R_{3D}$ which minimizes the scattering angles,
explicitly the following $\chi^2$ function:
\begin{equation}
	\chi^2(R_{3D}) = \frac{\Phi_{MS}(R_{3D})^2}{\sigma_\phi^2}  +
        \frac{\Theta_{MS}(R_{3D})^2}{\sigma_\vartheta^2} \ .
	\label{eq:chi1}
\end{equation}

For weak MS effects
the momentum dependence of the scattering uncertainty is negligible; the
case of large MS effects is discussed in more detail in section~\ref{sec:strong_MS}. 
Assuming $\frac{d \sigma_{MS}}{dR_{3D}}=0$, the minimization of $\chi^2(R_{3D})$ is thus equivalent to solving the equation
\begin{equation}
	\sin^2\vartheta \; \frac{d \Phi_{MS}(R_{3D})}{dR_{3D}} \;
    \Phi_{MS}(R_{3D})
    \; + \; \frac{d \Theta_{MS}(R_{3D})}{dR_{3D}} \;
    \Theta_{MS}(R_{3D}) = 0
	\label{eq:chi2}
\end{equation}
for $R_{3D}$. 
The scattering angle in the transverse plane $\Phi_{MS}$ is given by
\begin{equation}
	\Phi_{MS} = (\varphi_{12} - \varphi_{01}) - \frac{\Phi_1(R_{3D}) +
          \Phi_2(R_{3D})}{2} \ ,
\end{equation}
where the bending angles $\Phi_1$ and $\Phi_2$ are the solutions of
the transcendent equations
\begin{align}
	\sin^2 \frac{\Phi_1}{2} &= \frac{d_{01}^2}{4 R_{3D}^2} +
        \frac{z_{01}^2}{R_{3D}^2} \frac{\sin^2 \frac{\Phi_1}{2}}{\Phi_1^2} \ , 	\nonumber\\
	\sin^2 \frac{\Phi_2}{2} &= \frac{d_{12}^2}{4 R_{3D}^2} +
        \frac{z_{12}^2}{R_{3D}^2} \frac{\sin^2 \frac{\Phi_2}{2}}{\Phi_2^2} \ .  \label{eq:phieq}
\end{align}
These equations have several solutions depending on the number of
half-turns of the track. However, for most practical cases it is
sufficient to consider the first two solutions.

Similarly, the scattering angle in the longitudinal plane is given by
\begin{equation}
	\Theta_{MS} = \vartheta_2 -\vartheta_1 
\end{equation}
where the polar angles $\vartheta_1$ and $\vartheta_2$ can be calculated from the azimuthal bending angles using the relations
\begin{align}
 \sin \vartheta_1 &= \frac{d_{01}}{2 R_{3D}} \mathrm{cosec} \left(
 \frac{z_{01}}{2R_{3D} \cos \vartheta_1} \right) \ , \nonumber\\
 \sin \vartheta_2 &= \frac{d_{12}}{2 R_{3D}} \mathrm{cosec} \left(
 \frac{z_{12}}{2R_{3D} \cos \vartheta_2} \right) \label{eq:thetaeq} \ .
\end{align}
Alternatively the relations 
\begin{align}
	\Phi_1 &= \frac{z_{01}}{R_{3D}\cos \vartheta_1} \ , \nonumber\\
	\Phi_2 &= \frac{z_{12}}{R_{3D}\cos \vartheta_2} \label{eq:phitheta}
\end{align}
between the azimuthal bending angles and the polar angles can be
exploited.

Equations \ref{eq:phieq} and \ref{eq:thetaeq} have no algebraic
solutions; they can either be solved by numerical iteration or by
using a linearization around an approximate solution; the second
approach is discussed in the following.

\subsection{Taylor expansion around the circle solution}
The circle solution describes the case of constant curvature in the
plane transverse to the magnetic field $r_1 = r_2$ and no scattering
in that plane, $\Phi_{MS} = 0$.  This solution exists for any hit
triplet and is thus a good starting point for the linearization.  The
radius $R_C$ of the circle in the transverse plane going through three
points is given by
%
%
\begin{equation}
	R_C = \frac{d_{01} \; d_{12} \; d_{02}} {2 \; 
[(\mathbf{x_1}-\mathbf{x_0}) \times (\mathbf{x_2}-\mathbf{x_1})]_z},
\end{equation}
where $d_{ij}$ is the transverse distance between the hits $i$ and $j$
of the triplet, see Figure~\ref{fig:triplet}.

The bending angles for the circle solution are
\begin{align}
 \Phi_{1C} &= 2 \arcsin \frac{d_{01}}{2 R_C} \ , \nonumber\\
 \Phi_{2C} &= 2 \arcsin \frac{d_{12}}{2 R_C} \ .
 \label{eq:phic}
\end{align}
Note that the above equations have in general two solutions ($\Phi_{iC} < \pi$
and $\Phi_{iC} > \pi$) and care is needed to select the physical one,
especially for highly bent tracks.  
The corresponding three-dimensional radii of the arcs are calculated as
\begin{align}
	R_{3D,1C}^2 &= R_C^2 +\frac{z_{01}^2}{\Phi_{1C}^2} \ , \nonumber\\
	R_{3D,2C}^2 &= R_C^2 +\frac{z_{12}^2}{\Phi_{2C}^2}  \ .
\label{eq:r3dcircle}
\end{align}
In general $\Theta_{MS}\ne 0$ such that
the two radii are not identical. Using equation~\ref{eq:phitheta}, polar
angles for the circle solution are obtained:
\begin{align}
	\vartheta_{1C} &= \arccos \frac{z_{01}}{\Phi_{1C} R_{3D,1C}} \ , \nonumber\\
	\vartheta_{2C} &= \arccos \frac{z_{12}}{\Phi_{2C} R_{3D,2C}}  \ .
\label{eq:phithetacircle}
\end{align}
Starting from this special circle solution with no scattering in the
transverse plane, we calculate the general solution
$\Phi_{MS} \ne 0$ which minimizes equation~\ref{eq:chi1} and for
which momentum conservation is fulfilled, i.e.~$R_{3D}$ does not
change between the segments.  
With the positions of the three hits given, the arc lengths and the polar angles depend only
on the radius, i.e. $\Phi_{1,2}=\Phi_{1,2}(R_{3D})$ and
$\vartheta_{1,2}=\vartheta_{1,2}(R_{3D})$ (equations~\ref{eq:phieq}
and~\ref{eq:thetaeq}). 
We can therefore perform a Taylor expansion to first order
around the circle solution which is described by the parameters
$R_{3D,1C}$, $R_{3D,2C}$, $\Phi_{1C}$, $\Phi_{2C}$, $\vartheta_{1C}$
and $\vartheta_{2C}$:
\begin{align}
 \Phi_{1}(R_{3D}) &\approx  \Phi_{1C} + (R_{3D} - R_{3D,1C})\left.\frac{\d \Phi_1}{\d R_{3D}}\right|_{\Phi_{1C}}, \nonumber\\
 \Phi_{2}(R_{3D}) &\approx  \Phi_{2C} + (R_{3D} - R_{3D,2C})\left.\frac{\d \Phi_2}{\d R_{3D}}\right|_{\Phi_{2C}}
 \label{eq:phitaylor}
\end{align}
and
\begin{align}
 \vartheta_{1}(R_{3D}) &\approx  \vartheta_{1C} + (R_{3D} - R_{3D,1C})\left.\frac{\d \vartheta_1}{\d R_{3D}}\right|_{\vartheta_{1C}} \ , \nonumber\\
 \vartheta_{2}(R_{3D}) &\approx  \vartheta_{2C} + (R_{3D} - R_{3D,2C})\left.\frac{\d \vartheta_2}{\d R_{3D}}\right|_{\vartheta_{2C}} \ .
 \label{eq:thetataylor}
\end{align}

The derivatives $\left.\frac{\d \Phi_1}{\d R_{3D}}\right|_{\Phi_{1C}}$ and
$\left.\frac{\d \Phi_2}{\d R_{3D}}\right|_{\Phi_{2C}}$ can be represented by
index parameters:
\begin{align}
	\left.\frac{\d \Phi_1}{\d R_{3D}}\right|_{\Phi_{1C}} &= - \alpha_1 \frac{\Phi_{1C}}{R_{3D,1C}} \ , \nonumber \\
	\left.\frac{\d \Phi_2}{\d R_{3D}}\right|_{\Phi_{2C}} &= - \alpha_2 \frac{\Phi_{2C}}{R_{3D,2C}} \ ,
\label{eq:alphaderivatives}
\end{align}
which are calculated from equation~\ref{eq:phieq} as
\begin{align}
	\alpha_1 & = \frac{R_C^2 \Phi_{1C}^2 +z_{01}^2} {\frac{1}{2}R_C^2 \Phi_{1C}^3 \cot \frac{\Phi_{1C}}{2} + z_{01}^2} \ ,  \nonumber  \\
	\alpha_2 & = \frac{R_C^2 \Phi_{2C}^2 +z_{12}^2} {\frac{1}{2}R_C^2 \Phi_{2C}^3 \cot \frac{\Phi_{2C}}{2} + z_{12}^2} \ .
	\label{eq:alpha_index}
\end{align}
The derivatives of the polar angles are obtained from
equation~\ref{eq:thetaeq} and can be expressed by the same index parameters:
\begin{align}
	\left.\frac{\d \vartheta_1}{\d R_{3D}}\right|_{\vartheta_{1C}} = \frac{\cot \vartheta_{1C}}{R_{3D,1C}} (1-\alpha_1), \nonumber \\
	\left.\frac{\d \vartheta_2}{\d R_{3D}}\right|_{\vartheta_{2C}} = \frac{\cot \vartheta_{2C}}{R_{3D,2C}} (1-\alpha_2).
\end{align}

\subsection{Linearization of the scattering angles}
\label{sec:linearization}
The above relations can now be used to calculate a linearized
expression for the MS angles. For $\Phi_{MS}$ we
obtain:
\begin{align}
	\Phi_{MS} &= \varphi_{12} - \varphi_{01} - \frac{\Phi_1(R_{3D})}{2}  - \frac{\Phi_2(R_{3D})}{2} \nonumber \\
						&= \tilde{\Phi} + {\eta}\;
        R_{3D} \ ,
\end{align}
where we have introduced two new parameters:
\begin{align}
	\tilde{\Phi} &= - \frac{1}{2}(\Phi_{1C} \alpha_1 + \Phi_{2C} \alpha_2) \ ,\\
	{\eta} 			 &= \frac{\d\Phi_{MS}}{\d R_{3D}} = \frac{\Phi_{1C}\; \alpha_1}{2 R_{3D,1C}} + \frac{\Phi_{2C}\; \alpha_2}{2 R_{3D,2C}} \ .
\end{align}
And similarly for the polar angle $\Theta_{MS}$ we obtain:
\begin{align}
	\Theta_{MS} & = \vartheta_2 - \vartheta_1 \nonumber \\
							& = \tilde{\Theta} +
              {\beta}\; R_{3D} \ ,
\end{align}
with the new parameters
\begin{align}
	\tilde{\Theta}  & = \vartheta_{2C} - \vartheta_{1C} - \Bigl(\left(1-\alpha_2\right)\cot \vartheta_{2C} - \left(1-\alpha_1\right)\cot \vartheta_{1C}\Bigr) \ ,\\
	{\beta} &= \frac{\d\Theta_{MS}}{\d R_{3D}} = \frac{(1-\alpha_2)\cot
          \vartheta_{2C}}{R_{3D,2C}} - \frac{(1-\alpha_1)\cot
          \vartheta_{1C}}{R_{3D,1C}} \ .
\end{align}

\subsection{Linearized triplet track fit}

We can now minimize the $\chi^2$-function by inserting  the deri\-vatives and 
the expressions for the scattering angles obtained
from the linearization in equation~\ref{eq:chi2}.
For the three-dimensional radius we obtain
\begin{equation}
  R_{3D}^{min} = - \frac{{\eta}\;\tilde{\Phi}\;\sin^2\vartheta +
    {\beta}\;\tilde{\Theta}}{{\eta}^2\sin^2\vartheta + {\beta}^2} \ .
\label{eq:R3Dmin}
\end{equation}
Here is $\vartheta$ the polar angle at  the scattering layer, 
which can be taken as the 
average of $\vartheta_{1C}$ and $\vartheta_{2C}$.
The minimum $\chi^2$ value is
\begin{equation}
	\chi^{2}_{min} = \frac{1}{\sigma_{MS}^2}
    \frac{({\beta}\;\tilde{\Phi}
      - {\eta}\;\tilde{\Theta})^2}{{\eta}^2 + {\beta}^2/\sin^2\vartheta}
\label{eq:chi2min}
\end{equation}
and for the uncertainty of the three-dimensional radius we get
\begin{equation}
	\sigma(R_{3D}) = \sigma_{MS}\sqrt{\frac{1}{{\eta}^2\sin^2\vartheta +
            {\beta}^2}} \  .
\label{eq:sigmar3d}
\end{equation}
The scattering angles are finally given by:
\begin{align}
\Phi_{MS}  &=  {\beta} \ \frac{{\beta} \; \tilde{\Phi} - {\eta} \;
  \tilde{\Theta} }{{\eta}^2\sin^2\vartheta+{\beta}^2} \ , \\
\Theta_{MS}  &=  - {\eta} \sin^2\vartheta \ \frac{{\beta} \; \tilde{\Phi} -
  {\eta} \; \tilde{\Theta}}{{\eta}^2\sin^2\vartheta+{\beta}^2} \ .
\end{align}
It is straight-forward to calculate further track parameters using the
linearization described above.

Note that in this approach the fitted track parameters are independent of the
momentum and the MS uncertainty.
The latter can be calculated after fitting the track parameters
which allows for an elegant treatment of the material effects.
We have thus obtained a non-iterative solution to
the triplet problem with multiple
scattering.

\subsection{Strong Multiple Scattering and Weak Bending}
\label{sec:strong_MS}
The regime
where the MS uncertainty is of  similar size as
the sum of the bending angles, $\sigma_{MS} \approx \Phi_1+\Phi_2$,
we define as strong MS or weak bending.
This corresponds to cases with either a large amount of material 
at the scattering layer or weak magnetic field strength.
In this regime the momentum dependence of the
scattering uncertainty leads to a systematic shift (bias) of the fitted radius towards larger values.
This bias can be compensated by including the momentum dependence in the minimization
of the $\chi^2$ function, given in equation~\ref{eq:chi1}, using the ansatz
$\sigma_{MS} = b/R_{3D}$ which is motivated by the Highland
formula~\cite{Highland:1975:prm}. 
Here $b$ is
an effective scattering parameter which is approximately given by
\begin{equation}
b \approx \frac{4.5~\textrm{cm T}}{B} \; \sqrt{X/X_0} 
\end{equation}
and assumed to vary only weakly within the parameter range of the fit.  
The so obtained unbiased result  
\begin{equation}
R_{3D}^{unbiased}   =  - \; \frac{{\eta} \; \tilde{\Phi} \; \sin^2{\vartheta}  + {\beta}
  \; \tilde{\Theta}  }{{\eta}^2 \; \sin^2{\vartheta}  + {\beta}^2}
\  
\left(
\frac{3}{4} +  \; \frac{\sqrt{1-8 \; \delta^2  \sin^2{\vartheta} }}{4}
\right)
\label{eq:R3D_unbiased}
\end{equation}
with
\begin{equation}
\delta  =  \frac{{\beta} \; \tilde{\Phi} - {\eta} \; \tilde{\Theta}}
{{\eta} \; \tilde{\Phi} \; \sin^2{\vartheta}  + {\beta}
  \; \tilde{\Theta}  } 
\end{equation}
has only a solution if 
\begin{equation}
8 \; \delta^2 \sin^2{\vartheta} 
      \le 1 \ .
\label{eq:bias_free_condition1}
\end{equation}
For small bias parameters, $\delta \approx 0$, equation~\ref{eq:R3Dmin} is
restored. 
The bias parameter $\delta$ is just given by the hit triplet geometry but
it can also be expressed by fitted parameters:
\begin{equation}
\delta^2 \sin^2{\vartheta} 
\ = \ \frac{\sigma(R_{3D}^{min})^2}{{R_{3D}^{min}}^2} \; \chi^{2}_{min} 
\label{eq:delta}
\end{equation}
using equations~\ref{eq:R3Dmin}, \ref{eq:chi2min} and \ref{eq:sigmar3d}.
The bias term is thus proportional to the sum of the squared scattering angles:
\begin{equation}
\delta^2 \sin^2{\vartheta} 
\ \propto \   \Phi_{MS}^2 + \Theta_{MS}^2 \sin^2{\vartheta} \  .
\end{equation}

From equation~\ref{eq:bias_free_condition1} a  condition on the
minimum significance for the radius measurement can be derived:
\begin{equation}
	\frac{R_{3D}}{\sigma(R_{3D})} > \sqrt{8 \; \chi^2} \ .
\label{eq:R3Dsigni}
\end{equation}
If the radius significance is not large enough, 
$\frac{R_{3D}}{\sigma(R_{3D})} \lesssim 10$, significant 
bias corrections apply.

For small bending angles (weak bending region!) the relation
$|{\beta}| \ll |{\eta}|$ holds and 
the relative resolution of
the three-dimensional radius (momentum) is approximately given by
\begin{equation}
	\frac{\sigma(R_{3D})}{R_{3D}}= \frac{\sigma(p)}{p} = \frac{2 \;b}{s}
        \ .
\end{equation}
We can then rewrite equation~\ref{eq:R3Dsigni} as
\begin{equation}
s^2 > 32 \; b^2 \chi^2
	\label{eq:s}
\end{equation}
where $s=s_{01}+s_{12}$ defines the  length of the triplet trajectory.
This relation should be respected, for example in the design of detectors,
to allow for a decent momentum measurement.

\subsection{Example Spectrometer}

\begin{figure}[tb]
\centering
\includegraphics[width=0.5\textwidth]{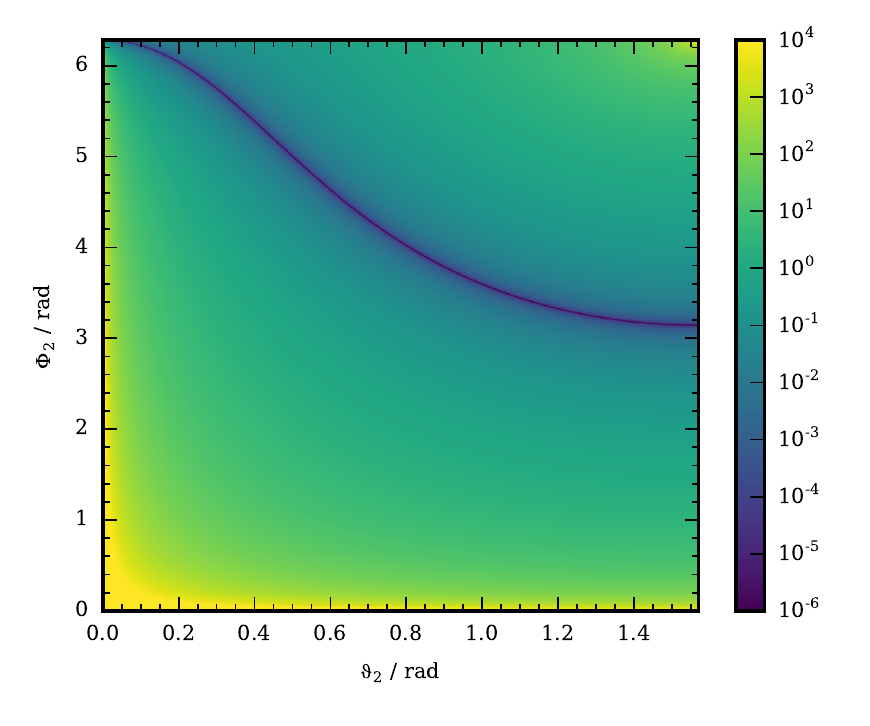}
\caption{Relative momentum resolution for the example spectrometer with
  the MS triplet fit, see equation \ref{eq:spectrometer}, as a
  function of the bending angle $\Phi_2$
  and the polar angle $\vartheta_2$.}
\label{fig:example_spectrometer_resolution}
\end{figure}

The resolution of the triplet track fit is investigated
for a simple spectrometer configuration with three detector layers
for which the spatial hit uncertainties are negligible. 
The first two layers are spaced closely
together and the third layer is placed further apart, i.e.~we assume
$\Phi_1 \ll \Phi_2$ for the sweep angles defined in \autoref{fig:triplet}.
The relative momentum
resolution is then calculated using the previously derived expressions
for the fitted radius and associated variances as follows:
\begin{equation}
\frac{\sigma_{R_{3D}}}{{R_{3D}}} =
2 \; \sigma_{MS} \; \left( \Phi_2^2 \alpha_2^2 \; \sin^2 \vartheta_2 \; + \; 4 (1-\alpha_2)^2 \;
    \cot^2 \vartheta_2 \right)^{-\frac{1}{2}},
\label{eq:spectrometer}
\end{equation}
with $\alpha_2$ given by:
\begin{eqnarray}
\alpha_2^{-1} & =& \cos^2{\vartheta_2} \; + \; \frac{\Phi_2}{2}
\cot{\left( \frac{\Phi_2}{2} \right)} \; \sin^2{\vartheta_2} \ .
\end{eqnarray}
The resulting resolution as a function of $\Phi_2$ and $\vartheta_2$ is shown 
in figure~\ref{fig:example_spectrometer_resolution}.
Note that for some special cases, if $1/\alpha_2 \rightarrow 0$,
the  momentum resolution approaches zero.
For transverse going tracks ($\vartheta_2 = \pi/2$)
this is the case if $\Phi_2 = \pi$ (i.e. semi-circles).
The geometry of track detectors in a regime where MS dominates
can thus be optimized for an almost perfect measurement at certain specific 
momenta.

\section{Combining Triplets}

\begin{figure}[t!]
  \centering
  \includegraphics[width=0.4\textwidth]{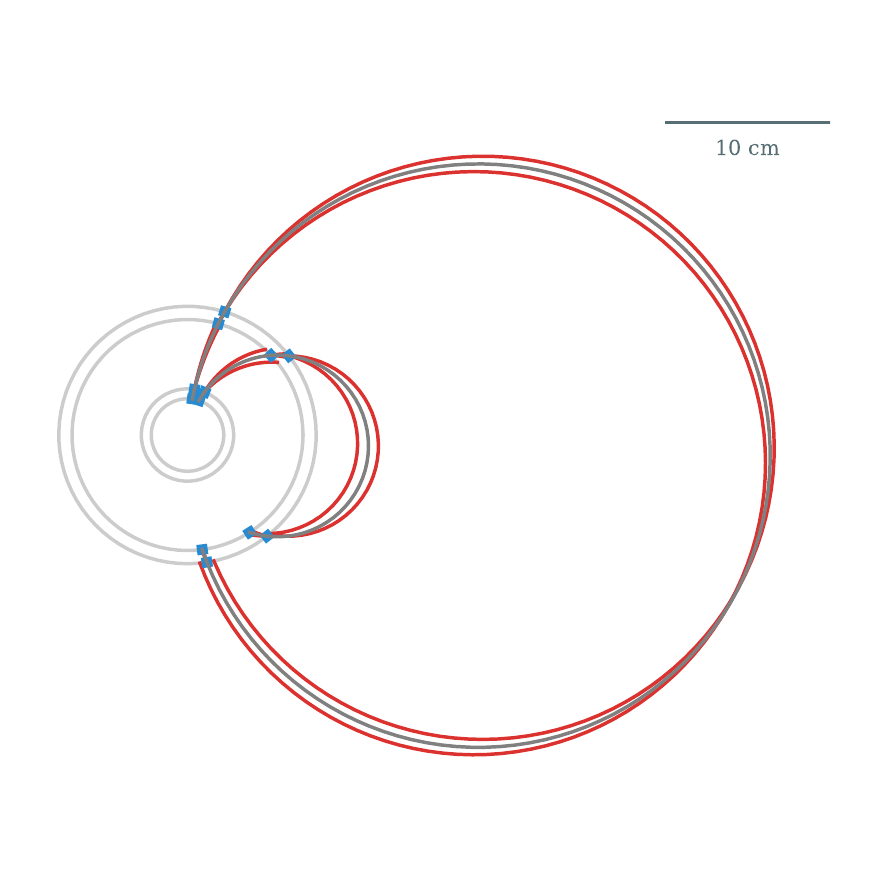}
  \caption{Simulated Mu3e tracker geometry in the transverse plane. 
    Two undisturbed example
    trajectories with a momentum of \SI{53}{\MeV\per{c}} and
    \SI{16.5}{\MeV\per{c}} and a fixed polar angle of
    $\vartheta = \SI{90}{\degree}$ are shown. 
    The red lines indicate the uncertainties induced by scattering in
    the previous layer for a particle moving outwards. The blue marks
    indicate the measurement uncertainties. All uncertainties are
    artificially increased by a factor five to enhance the
    visibility.  
    A magnetic field of
    \SI{1}{\tesla} is assumed.}
  \label{fig:mu3e-geometry}
\end{figure}
\begin{figure}[t!]
  \centering
  \includegraphics[width=0.45\textwidth]{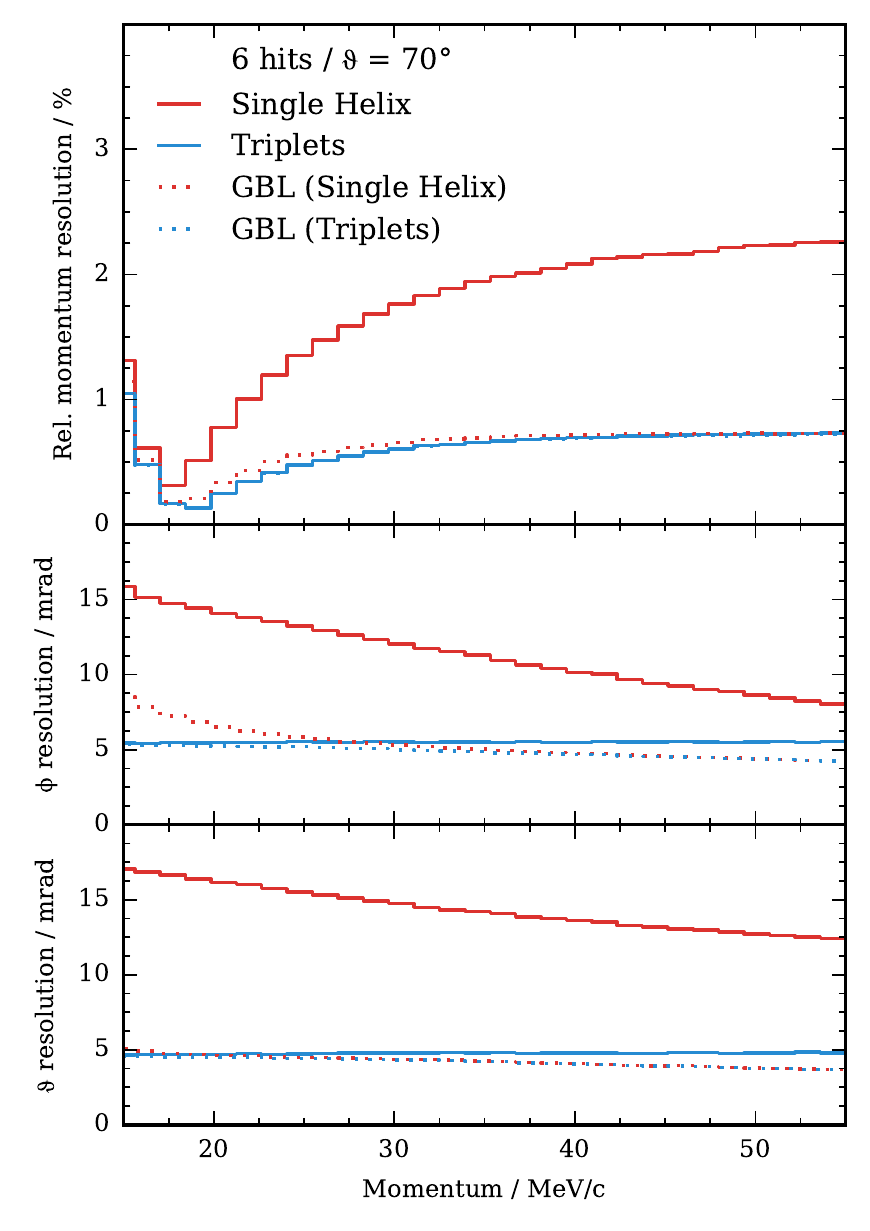}
  \caption{Track parameter resolution for the Mu3e geometry, described in
    \autoref{fig:mu3e-geometry}, for different track fits as a
    function of the track momentum. The top panel shows the relative
    momentum resolution and the bottom panels show the 
    resolution for the azimuthal angle $\phi$ and the polar angle
    $\vartheta$ for tracks with a polar
    angle of \SI{70}{\degree}. 
    The resolutions are calculated as the RMS of the
    parameter residual distributions for each bin.}
  \label{fig:mu3e-resolution}
\end{figure}

\begin{figure}[t!]
  \centering
  \includegraphics[width=0.4\textwidth]{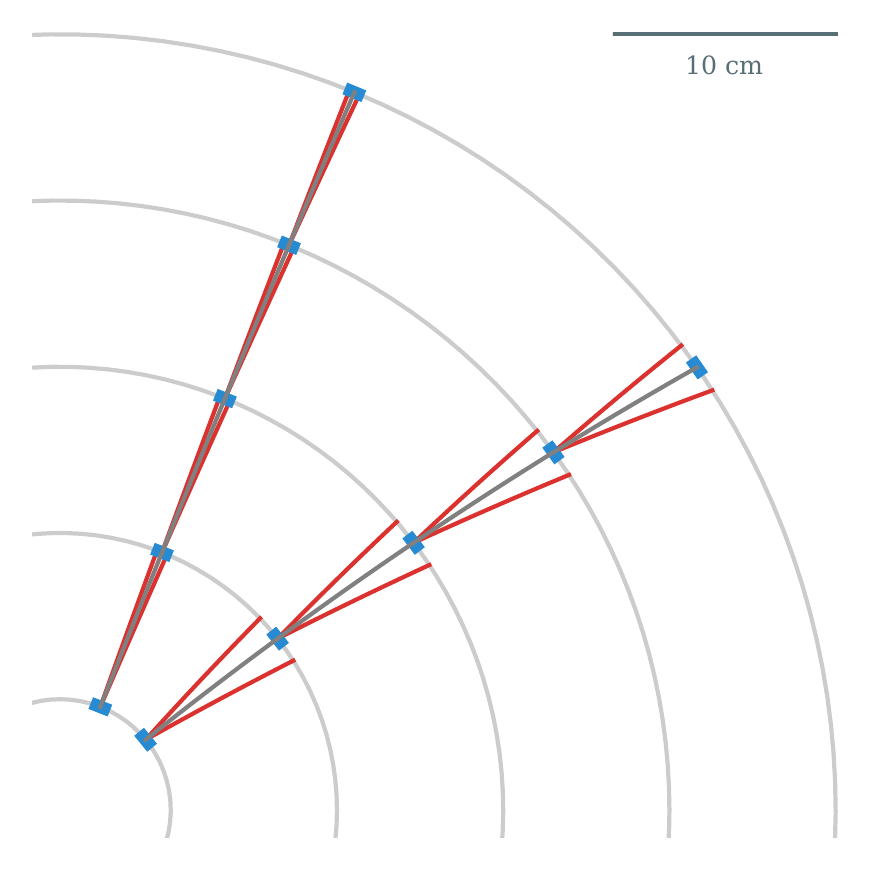}
  \caption{
    Simulated generic pixel tracker geometry with associated
    uncertainties in the transverse plane. Two undisturbed example
    trajectories are shown with momenta of \SI{5}{\GeV\per{c}} (upper)
    and \SI{1}{\GeV\per{c}} (lower) and a fixed polar angle of
    $\vartheta = \SI{70}{\degree}$.
    The uncertainties, see \autoref{fig:mu3e-geometry} for explanations,
    have a scaling factor of one-hundred. 
    A magnetic field of \SI{2}{\tesla} is assumed. }
  \label{fig:atlas_s5-geometry}
\end{figure}
\begin{figure}[t!]
  \centering
  \includegraphics{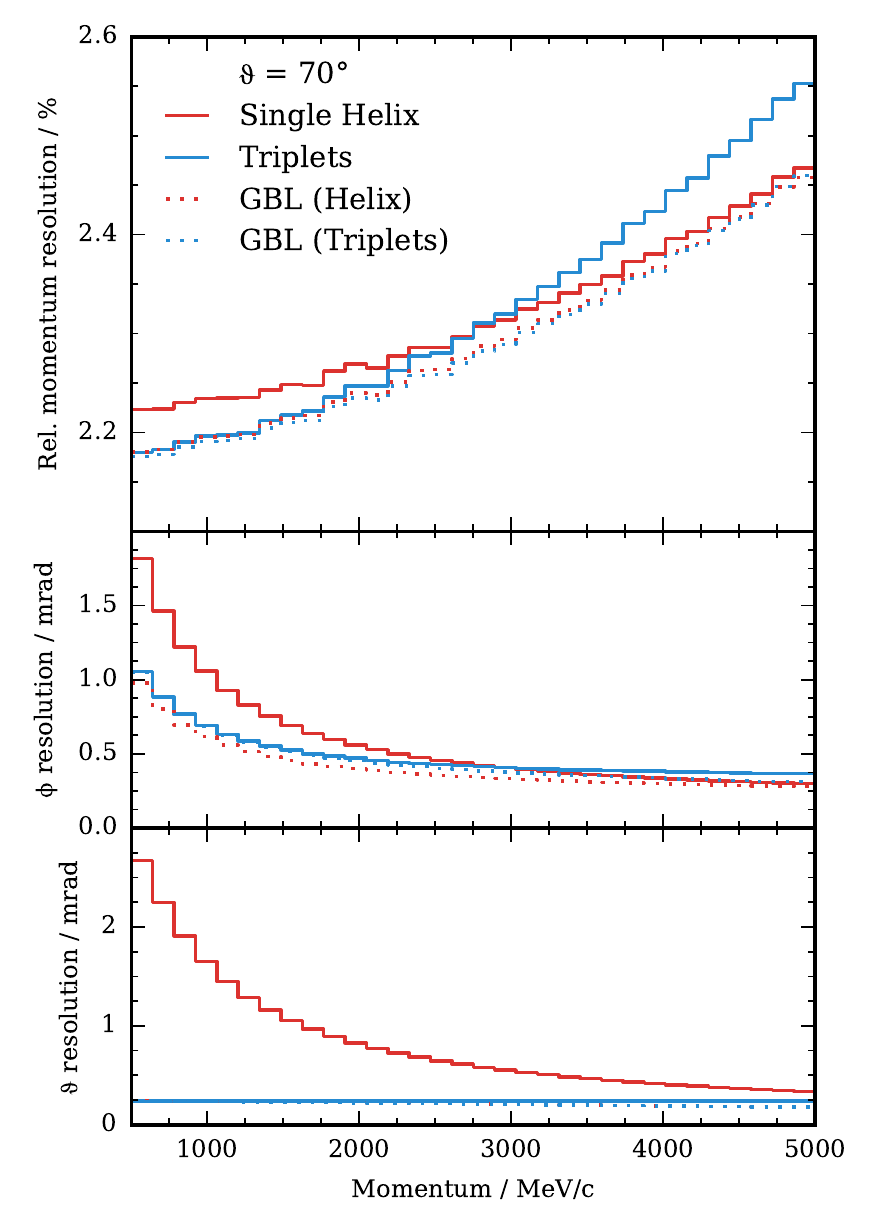}
  \caption{Track parameter resolution for the equidistant layer geometry,
    described in \autoref{fig:atlas_s5-geometry}, for different
    track fits as a function of the track momentum. The top panel shows
    the relative momentum resolution and the bottom panels show the
    resolution of the azimuthal angle $\phi$ and the polar
    angle $\vartheta$ for tracks with a
    polar angle of \SI{70}{\degree}. 
    The resolutions are calculated as the RMS of the
    parameter residual distributions for each bin.}
  \label{fig:atlas_s5-resolution}
\end{figure}

Several triplets can be combined to form longer tracks. 
As MS in each sensor layer is independent of all other layers the
following global $\chi^2$ function is minimized:
\begin{equation}
  \label{eq:chi2_global} 
  \chi_{global}^2 = \sum_{i}^{n_{hit}-2} \chi^2_i \ ,
\end{equation}
where $\chi_i$ is the minimization function for the $i$-th triplet
number previously defined in equation~\ref{eq:chi1}. The total number of triplets
is given by $n_{hit}-2$. The minimization of equation~\ref{eq:chi2_global}
is equivalent to a weighted average of the resulting radii
$R_{3D,i}$ of the individual fits:
\begin{equation}
	\overline{R_{3D}} = \sum_{i}^{n_{hit}-2}
        \frac{R_{3D,i}^3}{\sigma(R_{3D,i})^2} / \sum_{i}^{n_{hit}-2}
        \frac{R_{3D,i}^2}{\sigma(R_{3D,i})^2} \ .
\end{equation}
with the corresponding uncertainty given by
\begin{equation}
	\sigma(\overline{R_{3D}}) =
        \frac{\overline{R_{3D}}}{\sqrt{\sum_{i}^{n_{hit}-2}
            \frac{R_{3D,i}^2}{\sigma(R_{3D,i})^2}}} \ .
\end{equation}

This averaging formula is free of any bias if unbiased
three-dimensional radii from equation~\ref{eq:R3D_unbiased}
are used as input and if
$\frac{\sigma(R_{3D,i})}{R_{3D,i}}$ is constant, which is a very good
assumption for most cases.

Fitting of multiple hits, $n_{hit} > 3$, is performed
in a three step procedure: first all triplets of consecutive hits are
fitted individually, second a weighted mean of the three-dimensional
track radii is calculated which is then used in the third step to
recalculate all other track parameters.  It is worth to note that the
expected variance of the MS angle for each triplet only enters at
the averaging step, where it can be calculated to very good accuracy
from the locally fitted triplet track parameters.  Effects from
energy loss can also be incorporated at this step.

We can now use the average radius $\overline{R_{3D}}$ together
with our linearization to obtain globally fitted (updated) values for the
sweep angles $\Phi_i$ and the polar angles $\vartheta_i$:
\begin{align}
	\Phi_i^\prime &= \Phi_i -(\overline{R_{3D}} - R_{3D,i})
        \; \frac{\Phi_i}{R_{3D,i}}\; \alpha_i \ ,\\
	\vartheta_i^\prime  &= \vartheta_i - (\overline{R_{3D}} - R_{3D,i})
        \; \frac{\cot \vartheta_i}{R_{3D,i}} \; (1 - \alpha_i) \ .
\end{align}
Alternatively, $\vartheta_i^\prime$ can also be obtained from the relation
\begin{equation}
	\vartheta_i^\prime = \arccos \frac{{\eta} \;
          z_{\{i-1,i\}}}{\overline{R_{3D}}\; \Phi_i^\prime}
\end{equation}
if the sweep angle is known.

We have thus obtained a non-iterative solution to the MS
 problem which is especially suitable for implementation on
massively parallel architectures such as graphics processors (GPUs) as
the triplets can be fit in parallel.

\section{Track Fit Comparisons}

To compare the performance of the triplets fit with other fit algorithms we simulate
particle tracks in different detector geometries using a toy Monte Carlo.
Tracks are then reconstructed using the triplets fit, a single helix fit~\cite{Karimaki:1991:ecf}, and the
general broken lines (GBL) fit \cite{Kleinwort:2012:gbl}.

For the comparison study we choose two exemplary geometries.
Detector layers are modeled as cylindrical high resolution pixel
sensors centered around the origin and aligned along the direction 
of the homogeneous magnetic field. 
Tracks are generated at the origin 
and propagated in the magnetic field to the detector layers.  
MS is simulated by smearing the track
direction at each layer with kink angles drawn from a Gaussian
distribution with a width according to the Highland formula 
\cite{Highland:1975:prm, rpp2014}.
The position resolution of the detector is simulated by smearing
the registered hit positions with a Gaussian distribution along both
sensitive directions.

The triplets fit, which includes only MS uncertainties, 
is performed as described in the previous section. 

The single helix fit, which is another example for a direct track fit,
only takes into account the spatial measurement uncertainties.
Here, the transverse track parameters are obtained from the Karimäki circle fit 
\cite{Karimaki:1991:ecf} and the longitudinal parameters result from a linear
regression to the points in the projected arclength-$z$ plane.  

The GBL fit is an extended track fit that takes into account both,
scattering effects and spatial uncertainties.  It has been shown
\cite{Kleinwort:2012:gbl} to be equivalent to the \Kalman
filter \cite{Fruhwirth:1987:akf} and uses a track model consistent with all
simulated uncertainties. 

The GBL algorithm performs a linearized fit by varying positions and kink angles 
at selected points around a reference trajectory.
This reference can be derived from any direct track fit.  
Here, we use the helix fit and the triplets fit for comparison.
In the first case, the track parameters are refitted by
introducing non-vanishing kink angles.  In the second case, the kink 
angles are optimized by introducing 
residuals to the measured positions.
In this study we use only one iteration and  
ideally, the GBL converges to the same optimized trajectory from either of the 
two initial estimates in that single step.

For the comparison study, the fitted track parameters are calculated at the inner-most detector layer
at which the track parameters are maximally uncorrelated.

The first simulated configuration is the silicon pixel tracker of the
Mu3e experiment~\cite{BergerEtAl:2013:tme}. 
The geometry and example trajectories to
illustrate the uncertainties are shown in figure
\ref{fig:mu3e-geometry}. 
The detector is placed in a homogeneous magnetic field of $B=\SI{1}{T}$, 
has four layers with radii at
approximately \SIlist{2.2;2.8;7.0;7.8}{\cm}, and is optimized for low
momentum electrons in a momentum range of
\SIrange{15}{53}{\MeV\per{c}}. The spatial resolution on the detector
plane is $80/\sqrt{12}$~\SI{}{\um} and the layer thickness is \SI{0.1}{\percent}
radiation lengths.

The resulting parameter resolution for the four different fits is shown
in figure \ref{fig:mu3e-resolution}.  
The triplets fit has a consistently better resolution than the single helix fit
in all track parameters.
The GBL-fit with the triplets as reference (GBL-T) shows no significant
improvement at very small momentum ($\lesssim$\SI{25}{\MeV\per{c}}).
In this region MS uncertainties dominate the track
uncertainties. Above \SI{25}{\MeV\per{c}} the GBL-T fit allows for small
improvements of the angular resolutions as spatial uncertainties start to
contribute.

The GBL-fit with the single helix reference (GBL-H) reproduces for
$\gtrsim$\SI{30}{\MeV\per{c}} the result of GBL-T.
Although GBL-H improves with respect to the single helix reference fit also
for $\lesssim$\SI{30}{\MeV\per{c}}, the momentum and azimuthal angle
resolution is worse than for triplets fit and GBL-T.
In this region the single helix parameterization is not a good reference and
linearization point for the GBL. 
With the large non-linearity of the strongly curved tracks the single GBL-H
step is insufficient.

The behaviour of the position resolution (not shown) follows the behaviour of the
angle resolution, i.e.~the GBL is slight\-ly better than the triplets fit
and the helix fit is significantly worse than other fits.

To extend this study beyond the unique Mu3e configuration, a generic
pixel tracker design similar to existing or planned trackers for high-energy
collider detectors, e.g. ATLAS, CMS or ILC, is evaluated.  
It comprises five equidistant detector layers at radii between 
\SIrange{40}{340}{\mm} with a spatial
resolution of 
$50/\sqrt{12}$~\SI{}{\um} and a sensor thickness corresponding to
\SI{2}{\percent} of radiation
length. 
Particles are simulated in the  momentum range between
\SIrange{500}{5000}{\MeV\per{c}}, a region where MS
significantly contributes to the track uncertainties. 

The track parameter resolution of the four algorithms is shown in 
figure~\ref{fig:atlas_s5-resolution}. 
All fits show a similar momentum resolution. 
At low momentum the triplets fit provides the better resolution, at higher
momentum the single helix fit, with a crossover at around
\SI{3000}{\MeV\per{c}}.
The GBL-fits give the optimal resolution over the full range.

The polar angle resolution of the triplets fit is constant, its value fully
determined by the spatial hit uncertainties.
Interestingly, the triplets fit gives a significantly better resolution than
the single helix fit even at high momenta.
For the GBL-fits an improvement of the  polar angle resolution 
for $\gtrsim$\SI{3000}{\MeV\per{c}} is visible.

The azimuthal angle resolution shows a similar crossover behaviour as the
momentum resolution at about \SI{3000}{\MeV\per{c}}. 
Again both GBL-fits lead to an improvement of the resolution.
But they show a small difference at low momentum. 
Interestingly, the first iteration step of GBL-H yields  a better
azimuthal resolution than GBL-T.\footnote{After sufficient iteration the GBL-H resolution
  converges to the GBL-T resolution, which in turn agrees with the resolution of the
  triplets fit.}
Note that the position of the crossover point
depends on the geometry, the material, and the
spatial resolution of the detector.  

We have compared execution times and the number of floating point operations for
several implementations of the triplets fit and the single helix fit.
The number of cycles required varies greatly depending on how many geometric
quantities are pre-calculated and cached and whether (and where) a covariance
matrix is calculated.
With ideal caching and no calculation of the covariance matrix, 
the triplets fit outperforms the
single helix fit by almost a factor of 2; if all track parameters 
and the full covariance matrix is calculated at each hit position,
the single helix fit (with its global covariance matrix) needs about a factor 2 (5)
less cycles for three (eight) hits.

\section{Conclusions}

We presented a new track fit algorithm, the triplets fit, that is only based on
MS uncertainties to determine global momentum and local
direction parameters.
The triplets fit is motivated by the excellent position resolution 
of modern silicon pixel sensors which create track fitting problems
with dominating MS uncertainties.

Although developed initially for reconstructing very low momentum
electrons in the Mu3e experiment the triplets fit exhibits good performance for
pixel trackers at the high energy experiments at LHC where MS
uncertainties dominate or significantly contribute up to around \SI{10}{\GeV\per{c}}.
In this regime, the performance of the triplets fit
is as good as for GBL-fits.

The triplets fit enables a very fast computation of track parameters 
and provides a natural scheme for track finding and linking via the
combination of single triplets. 
This makes the triplets fit ideally suited for fast online reconstruction,
as a reference for extended track fits, and as fast algorithm for 
pattern recognition problems.

\section*{Acknowledgements}

N.~Berger and A.~Kozlinskiy would like to thank the Deut\-sche
Forschungsgemeinschaft for support through an Emmy Noether
grant and the PRISMA cluster of excellence at Johannes Gutenberg University Mainz. 
M.~Kiehn acknowledges support by the International Max Planck
Research School for Precision Tests of Fundamental symmetries.

\bibliography{references}

\begin{thebibliography}{10}
\expandafter\ifx\csname url\endcsname\relax
  \def\url#1{\texttt{#1}}\fi
\expandafter\ifx\csname urlprefix\endcsname\relax\def\urlprefix{URL }\fi
\expandafter\ifx\csname href\endcsname\relax
  \def\href#1#2{#2} \def\path#1{#1}\fi

\bibitem{Karimaki:1991:ecf}
V.~Karim{\"a}ki,
  \href{http://www.sciencedirect.com/science/article/pii/016890029190533V}{Effective
  circle fitting for particle trajectories}, Nucl. Instr. Meth. A 305~(1)
  (1991) 187--191.
\newblock \href {http://dx.doi.org/10.1016/0168-9002(91)90533-V}
  {\path{doi:10.1016/0168-9002(91)90533-V}}.
\newline\urlprefix\url{http://www.sciencedirect.com/science/article/pii/016890029190533V}

\bibitem{Kalman:1960:nal}
R.~E. Kalman,
  \href{http://FluidsEngineering.asmedigitalcollection.asme.org/article.aspx?articleid=1430402}{A
  new approach to linear filtering and prediction problems}, J. Basic Eng.
  82~(1) (1960) 35.
\newblock \href {http://dx.doi.org/10.1115/1.3662552}
  {\path{doi:10.1115/1.3662552}}.
\newline\urlprefix\url{http://FluidsEngineering.asmedigitalcollection.asme.org/article.aspx?articleid=1430402}

\bibitem{Fruhwirth:1987:akf}
R.~Fr{\"u}hwirth,
  \href{http://www.sciencedirect.com/science/article/pii/0168900287908874}{{Application
  of Kalman filtering to track and vertex fitting}}, Nucl. Instr. Meth. A
  262~(2{\textendash}3) (1987) 444--450.
\newblock \href {http://dx.doi.org/10.1016/0168-9002(87)90887-4}
  {\path{doi:10.1016/0168-9002(87)90887-4}}.
\newline\urlprefix\url{http://www.sciencedirect.com/science/article/pii/0168900287908874}

\bibitem{BilloirEtAl:1990:spr}
P.~Billoir, S.~Qian,
  \href{http://www.sciencedirect.com/science/article/pii/016890029091835Y}{{Simultaneous
  pattern recognition and track fitting by the Kalman filtering method}}, Nucl.
  Instr. Meth. A 294~(1{\textendash}2) (1990) 219--228.
\newblock \href {http://dx.doi.org/10.1016/0168-9002(90)91835-Y}
  {\path{doi:10.1016/0168-9002(90)91835-Y}}.
\newline\urlprefix\url{http://www.sciencedirect.com/science/article/pii/016890029091835Y}

\bibitem{Blobel:2006:nft}
V.~Blobel, A new fast track-fit algorithm based on broken lines, Nucl. Instr.
  Meth. A 566 (2006) 14--17.

\bibitem{BlobelEtAl:2011:fac}
V.~Blobel, C.~Kleinwort, F.~Meier,
  \href{http://www.sciencedirect.com/science/article/pii/S0010465511001093}{Fast
  alignment of a complex tracking detector using advanced track models},
  Comput. Phys. Commun. 182~(9) (2011) 1760--1763.
\newblock \href {http://dx.doi.org/10.1016/j.cpc.2011.03.017}
  {\path{doi:10.1016/j.cpc.2011.03.017}}.
\newline\urlprefix\url{http://www.sciencedirect.com/science/article/pii/S0010465511001093}

\bibitem{Kleinwort:2012:gbl}
C.~Kleinwort,
  \href{http://www.sciencedirect.com/science/article/pii/S0168900212000642}{General
  broken lines as advanced track fitting method}, Nucl. Instr. Meth. A 673
  (2012) 107--110.
\newblock \href {http://dx.doi.org/10.1016/j.nima.2012.01.024}
  {\path{doi:10.1016/j.nima.2012.01.024}}.
\newline\urlprefix\url{http://www.sciencedirect.com/science/article/pii/S0168900212000642}

\bibitem{BergerEtAl:2013:tme}
N.~Berger, et~al.,
  \href{http://www.sciencedirect.com/science/article/pii/S016890021300613X}{{A
  tracker for the Mu3e experiment based on high-voltage monolithic active pixel
  sensors}}, Nucl. Instr. Meth. A 732 (2013) 61--65.
\newblock \href {http://dx.doi.org/10.1016/j.nima.2013.05.035}
  {\path{doi:10.1016/j.nima.2013.05.035}}.
\newline\urlprefix\url{http://www.sciencedirect.com/science/article/pii/S016890021300613X}

\bibitem{mu3e-rp}
A.~Blondel, et~al., \href{http://arxiv.org/abs/1301.6113}{Research proposal for
  an experiment to search for the decay \{\ensuremath{\backslash}mu\}
  -\ensuremath{>} eee}, {ArXiv13016113} Hep-Ex Physicsphysics.
\newline\urlprefix\url{http://arxiv.org/abs/1301.6113}

\bibitem{Highland:1975:prm}
V.~L. Highland,
  \href{http://www.sciencedirect.com/science/article/pii/0029554X75907430}{Some
  practical remarks on multiple scattering}, Nucl. Instr. Meth. A 129~(2)
  (1975) 497--499.
\newblock \href {http://dx.doi.org/10.1016/0029-554X(75)90743-0}
  {\path{doi:10.1016/0029-554X(75)90743-0}}.
\newline\urlprefix\url{http://www.sciencedirect.com/science/article/pii/0029554X75907430}

\bibitem{rpp2014}
K.~A. Olive, et~al.,
  \href{http://inspirehep.net/record/1315584/export/hx}{Review of particle
  physics}, Chin.Phys. C38 (2014) 090001.
\newblock \href
  {http://dx.doi.org/10.1088/1674-1137/38/9/090001,\%002010.1088/1674-1137/38/9/090001}
  {\path{doi:10.1088/1674-1137/38/9/090001,\%002010.1088/1674-1137/38/9/090001}}.
\newline\urlprefix\url{http://inspirehep.net/record/1315584/export/hx}

\end{thebibliography}

\end{document}